\documentclass[12pt,twoside,pointlessnumbers,nofootinbib,smallheadings]{revtex4}
\usepackage{amsmath}
\usepackage{amssymb}
\usepackage{color}

\usepackage{axodraw4j}

\usepackage{hangcaption,fancybox,feynmp}
\usepackage{indentfirst}
\usepackage{graphicx}
\usepackage{epsfig,subfigure,psfrag}
\usepackage{CJK}
\usepackage{mathrsfs}

\begin{document}

%%%%%%%%%%%%%%%%%%%%%%%%%%%%%%%%%%%

\title{
Four Jets as a Probe of O(100 GeV) Physics beyond Standard Model at Hadron Colliders
}
\author{Xiao-ping Wang $^{1}$ \footnote{E-mail:hcwangxiaoping@pku.edu.cn},
Jia Xu $^{1}$ \footnote{E-mail:jiaxupku@pku.edu.cn},
 and Shou-hua
Zhu $^{1,2}$ }

\affiliation{
$ ^1$ Institute of Theoretical Physics $\&$ State Key Laboratory of
Nuclear Physics and Technology, Peking University, Beijing 100871,
China \\
$ ^2$ Center for High Energy Physics, Peking University,
Beijing 100871, China }

\date{\today}

\maketitle

\begin{center}
{\bf Abstract}

\begin{minipage}{15cm}
{\small  \hskip 0.25cm

O(100 GeV) physics beyond the standard model (BSM) could be overlooked provided that it is hidden in the untouched Higgs sector or
jets. The top quark forward-backward asymmetry measurements and di-jet bump, which is observed in the associated production with charged
lepton plus missing energy (supposed arising from W decay), may indicate the existence of a new color-octet axial-vector $Z_C$ with a mass
about 145 GeV. Here $Z_C$ decays into two jets with a certain branching ratio. In this paper we investigated the possibility to discover $Z_C$ pair via analyzing the four jets as the final states,
which are heavily polluted by huge QCD background. Our simulation
showed that, however, both Tevatron and LHC have the excellent chance to discover $Z_C$ through analyzing the four jets events
with accessible integrated luminosity and good control of QCD background.

 }

\end{minipage}
\end{center}

%\renewcommand{\baselinestretch}{1.2}
%\fontsize{12pt}{12pt}\selectfont

\newpage
%------------------------denotation------------
%Red: change
%Blue: From Bhattacherjee et.al.
%Yellow: From
%----------------------------------------------
\section{Introduction\label{Introduction}}
The Standard Model (SM) of particle physics has been extensively
tested at the LEP, Tevatron, and current running LHC with $\sqrt{s}=7$
TeV. Even though several anomalies between theoretical prediction
and experimental measurement persisted, such as the forward-backward asymmetry of
b-quark pair production at LEP
\cite{Tricomi:2002bb,Abbaneo:1998xt},
the forward-backward asymmetry of
top-quark pair production at Tevatron
\cite{Abazov:2007qb,Aaltonen:2008hc,CDFnote9724,D0Note6062.CONF,
CDFnote10224},
and the di-jet anomalous
bump debated between CDF and D0 of Tevatron
\cite{Aaltonen:2009vh,Aaltonen:2011mk}, most of measurements
are still consistent with the SM predictions and no obvious
deviations have been discovered yet.

Why is the SM so successful? As usually conjectured, provided
that the physics beyond the SM (BSM) does exist, BSM must be at O(TeV) scale or even higher.
Discovering such
kind of BSM is one of the main goals of high energy LHC. Another possible answer
to the question is that
the scale of BSM is not so high but hidden
in the untouched sector of the SM, the Higgs sector, or the not-so-well
measured objects, namely jets. The first case is due to the small signal rate and the
latter one is due to the huge QCD background, especially when the BSM scale is at O(100 GeV). In this paper,
we will pursue the second possibility and concentrate on the BSM with one
new color-octet vector boson $Z_C$, which decays into two jets with a certain branching ratio. We should emphasize that the study on $Z_C$ is just
for the purpose of demonstration. The multi-jet events can be acted as the probe, in the much broader
sense, to investigate the BSM at O(100 GeV) at the Tevatron and the LHC.

The new color-octet vector boson $Z_C$ was initially proposed \cite{Xiao:2010ph,Wang:2011hc} to account for the
the forward-backward asymmetry of
top-quark pair production at Tevatron.
The forward-backward asymmetry obtains the required contribution from the
interference among the s-channel $Z_C$-mediating diagram and the usual QCD ones. At the same time, if the $Z_C$ couples with
the quarks axial-like, the theoretical predictions are still consistent with the
experimental measurements of the differential and total cross section of top pair production.
Later, $Z_C$ was utilized to explain the anomalous di-jet bump discovered by
CDF  \cite{wang:2011taa}. The di-jet is associated production with $W$. It should be noted that such di-jet bump is not confirmed by D0\cite{Abazov:2011af}. However it is
quite interesting to search such di-jet bump at LHC and at Tevatron with more accumulated data (roughly $10fb^{-1}$) and investigate
the different capacities to isolate the different mechanism to produce such  di-jet bump. For example at LHC the main contributions may
come from gluon-gluon contributions while at Tevatron the quark contributions are larger. In a sense Tevatron can be a better
collider to observe the multi-jet signal arising from quark annihilation. This is one of the reasons why we still keep on focusing on the detail
searching strategies of multi-jet events at Tevatron.

The new color-octet particle can be potentially discovered via di-jet measurement. However due to huge QCD background, the di-jet is not
always an excellent final states to probe new particle. For example at Tevatron the constraint for particle with mass less than 200 GeV is weak
\cite{Aaltonen:2008dn}.
However the earlier and less energetic experiments at UA2 can
provide the strictest constraints \cite{Smith:1987fv,Sumorok:1985eb,Lubrano:1990et,Pastore:1989aj,Alitti:1993pn},
as also shown in Refs.
\cite{wang:2011taa,Wang:2011uq}.
Based on above observations,
pair production of new color-octet particle with mass less than 200 GeV can be a useful mode. Provided that a new particle decays into di-jet, the pair
production will induce four jets. With more jets one can utilize new handles, which depend on the characteristics of physics to be investigated,
 to suppress corresponding QCD background. To investigate how to suppress the
huge QCD background is one of the main topics in this paper. In Ref.
\cite{Dobrescu:2007yp}, the authors have investigated the
possibility of observing four jets from massive color-octet scalar
boson pair at the
Tevatron/LHC\cite{Kilic:2008ub}. Experimentally,
at Tevatron there were several analysis of six jets events
 in order to find low mass squarks and gluinos \cite{Aaltonen:2011sg,Aaltonen:2008rv,:2007ww}.
Recently
ATLAS \cite{Aad:2011yh} released
measurements for low mass pair-produced color-octet scalar particles, which subsequently decaying into the
four-jet final state. Due to the low integrated luminosity, the constraints from ATLAS are still weak. Based on
the measurements from Ref. \cite{Aad:2011yh}, we
will set limit for our model parameters after considering the different acceptances for pair production of
scalar and vector particles.

The paper is organized as following. In section II, we describe briefly the coupling among $Z_C$ and quarks as well as
gluon. The phenomenology of $Z_C$ is also investigated in this section. In Section III, we simulate the signal and background of the four-jet events at Tevatron and
LHC with $\sqrt{s}=7$ TeV and $14$ TeV respectively and
study the useful cuts to suppress the QCD background. Section IV contains our conclusions and discussions.

\section{$Z_C$ and its phenomenology}

\subsection{Interactions of $Z_C$}

The new color-octet vector can easily appear if one extends the
color gauge group $SU(3)_C$ to the larger group, for example
$SU(3)_1 \otimes SU(3)_2$ with the gauge couplings $h_1$ and $h_2$
respectively. Introducing scalar field $\Phi(3,\bar{3})$ under
$SU(3)_1 \otimes SU(3)_2$ with vacuum expectation $M\times
I_{3\times3}$, we could break down $SU(3)_1 \otimes SU(3)_2$ to the usual $SU(3)_C$.

\begin{figure}[htbp]
 \begin{center}
  \includegraphics[width=0.8\textwidth]{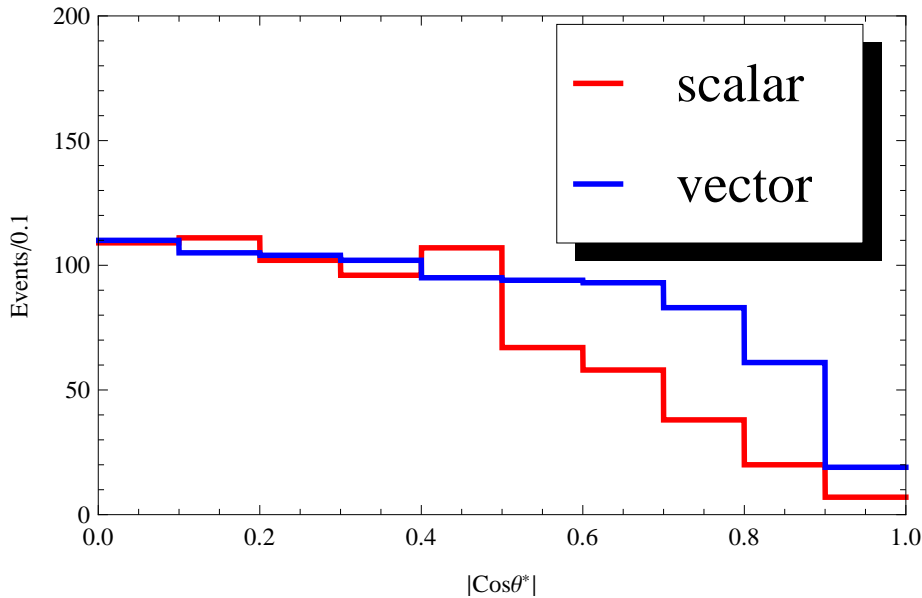}
  \end{center}
  \caption{Distributions of scalar-octet and vector-octet as the function of $\cos\theta^*$ with the same cross section after applying
  $p_T$, $\Delta R$, and mass difference cuts. The detailed description can be found in Ref. \cite{Aad:2011yh}. }
  \label{sov}
\end{figure}

The interactions of $Z_C$ and gluons can be written as
\begin{eqnarray}
&&    \frac{1}{2}{g_s}^2f^{abc}f^{ade}Z_C^{\mu\ b}\left[G^{\nu\ d}(Z_{C\nu}^{c}
G^{e}_{\mu}+Z_{C\mu}^{e}G^{c}_{\nu})+Z_{C\nu}^{e} G^{\nu\
c}G^{d}_{\mu}\right]
 \nonumber\\
&&  +g_{s}f^{abc}Z^a_{C\mu}\left[(\partial^{\mu}Z_C^{\nu\
b}-\partial^{\nu}Z_C^{\mu\ b})G^c_{\nu}-Z_{C\nu}^{b}
\partial^{\mu}G^{\nu\ c}\right]
\end{eqnarray}
Here $g_s$ is the usual strong coupling constant, $f^{abc}$ are the
$SU(3)$ structure constants, $G_{\mu}$ is the gluon field.
The color-octet new particles have been searched by experiment.
Recently ATLAS \cite{Aad:2011yh} has measured the four-jet cross section. From
their measurement the limit of the cross section times the branching
ratio to four-jet final states for color-octet
scalar \cite{Choi:2008ub,Schumann:2011ji} is roughly 1nb when the
mass is $145GeV$. In order to deduce the limit of the vector $Z_C$,
we simulate the scalar and vector at the same time and apply the
same cuts as those of ATLAS experiment\cite{Aad:2011yh}. Furthermore, in order to compare the difference between
scalar-octet and vector-octet, we illustrate the $\cos\theta^*$
distribution in Fig \ref{sov}. Theoretically $\theta^*$ is the open angle between $Z_C$ and beam in the center-of-mass frame of
final states.
We find that in the signal region ($|\cos(\theta^*)|<0.5$) the vector-octet's contribution is almost the same as scalar-octet, and in the background
region ($|\cos(\theta^*)|>0.7$), although the vector-octet' contribution is larger, they are in the same order. Based on this,
the limit cross section times the branching ratio to
four-jet final states for the vector-octet is safely to be the same as the scalar-octet', which is $1nb$ at 145 GeV.
Then we set the branching ratio to four-jet final states to be
$Br^2=0.06$, which $Br$ denotes $Br(Z_{C}\rightarrow jj)$.
The branching ratio of $Z_{C}$ into two-jet final states naturally
arise from many complete model construction to guarantee the guage
anomaly cancelation by introducing exotic states. The nature of
these exotic particles is model dependent. In other words, the
branching ratio of $Z_{C}$ into two-jet final states is a free
parameter.

The effective Lagrangian of $Z_C$ and quarks can be written as
\begin{eqnarray}
\Delta\mathcal{L}_q=\beta g_s Z_C^{\mu a}J^{5 a }_{\mu},
\label{beta}
\end{eqnarray}
where $J^{5 a }_{\mu}$ is the axial vector color current
\begin{eqnarray}
J^{ 5 a}_{\mu}=\sum_{f}\bar{q}_f\gamma_{\mu} \gamma_5 \frac{\lambda^a}{2}q_f
\end{eqnarray}
with $\lambda^a$ the Gell-Mann color matrices. Such axial-vector
current can account for the top quark forward-backward asymmetry and
the di-jet anomalous bump simultaneously
\cite{Xiao:2010ph,Wang:2011hc,wang:2011taa}, although by introducing the branching ratio of $Z_C$ into two jets, the di-jet bump's significance will decrease. To evade the limit from
the di-jet measurements by UA2, similar to the case in Ref.
\cite{wang:2011taa}, $\beta$ will be less than 0.3, which could be
obtained according to the model construction in
\cite{Dobrescu:2007yp}. In our following numerical evaluation, we
set $\beta$ to be $0.2 $.

\subsection{$Z_C$ Phenomenology}

As the color-octet particle, $Z_C$ can decay only into quarks, i.e. jets, if kinematically allowed. The main production mechanism at hadron colliders
is the single $Z_C$ production via $q\bar q \rightarrow Z_C$. However, as emphasized in Introduction, for $m_{Z_C} <200$ GeV, the QCD backgrounds at Tevatron
and LHC are overwhelming.

$Z_C$ can be associated production with $\gamma, W$ and $Z$. In fact the $WZ_C$ production mode is assumed to be the origin of
di-jet anomalous bump observed by CDF. In Ref.  \cite{wang:2011taa}, we have investigated this channel. Our study showed that  $WZ_C$ mode
can act as the discovery channel with low integrated luminosity. $Z_C$ can be associated production with light quarks and gluons and the final states
will be three jets. One can expect such signal is very likely buried  by huge QCD backgrounds, similar to the case of di-jet. In this paper we
will focus on the $Z_C$ pair production and the final states are four jets. Though the QCD background is still very large, we can find ways to suppress them greatly,
as shown below.

\begin{figure}[htbp]
 \begin{center}
  \includegraphics[width=0.8\textwidth]{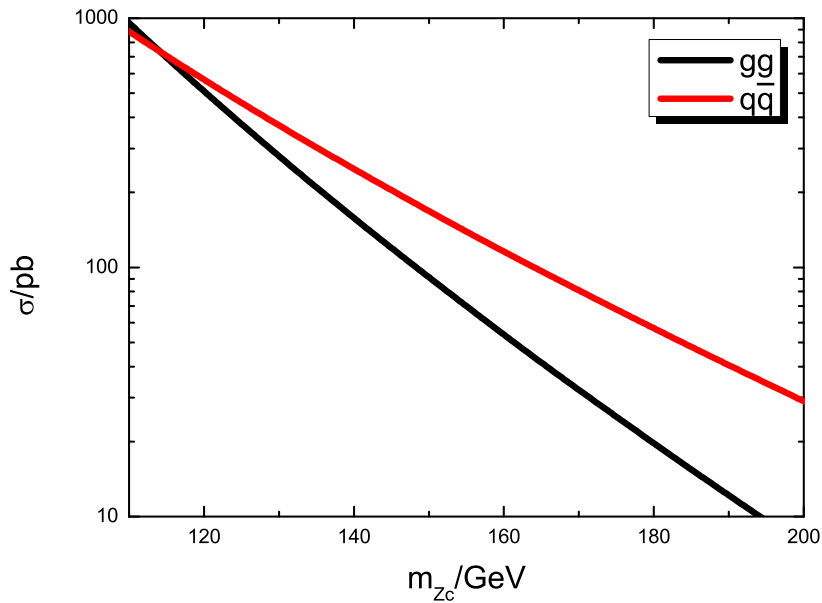}
  \end{center}
  \caption{The cross sections of $Z_C$ pair production as a function of $m_{Z_C}$ induced by $q\bar{q}$ and gluon-gluon sub-processes at Tevatron with the parameter introduced in Eq.\ref{beta}}
  \label{zcpairtev}
\end{figure}

\begin{figure}[htbp]
 \begin{center}
  \includegraphics[width=0.8\textwidth]{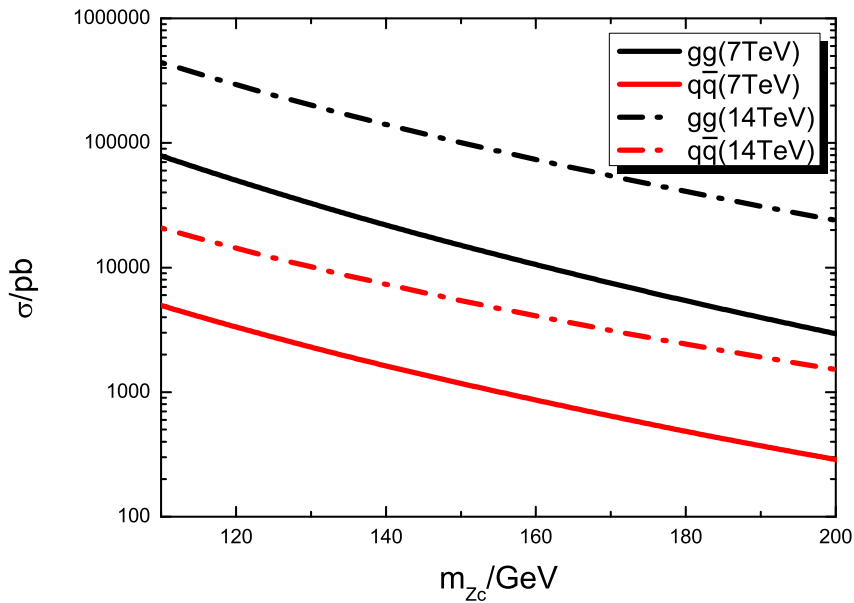}
  \end{center}
  \caption{Same with Fig. \ref{zcpairtev} except at LHC with $\sqrt{s}=7$ and $14$ TeV respectively.  }
  \label{zcpairlhc}
\end{figure}

The signal cross sections of $q\bar q (gg) \rightarrow Z_C Z_C$ as a function of $m_{Z_C}$ are depicted in Fig. \ref{zcpairtev} at
Tevatron and  Fig. \ref{zcpairlhc} at LHC. From the curves it is obvious that the $Z_C$ pair production is mainly from $q\bar q$ sub-processes at Tevatron while
from gluon-gluon fusion at LHC. Thus Tevatron and LHC play the different role to identify the different production mechanisms. The signal cross sections at Tevatron
and LHC are both large. Provided that the QCD backgrounds can be controlled, the signal will be easily identified. This will be the topic of next section.

\section{Detailed simulation}

In order to suppress the huge QCD background, we will study the different behaviors of signal and background and find the
optimized cuts to isolate the signal. In the simulation, $m_{Z_C}$ is set to be 145 GeV.
We use $MadGraph$
\cite{Alwall:2011uj,Maltoni:2002qb} to generate four jets events arising from signal and QCD background.
The resulting events are put into $Pythia$ and $PGS$ packages to do
fragmentation, hadronization and detector simulation.

\subsection{$Z_C$ at Tevatron}

The basic cut are chosen as
\begin{equation}
p_{T,j}\geq 50GeV;\ \ \ \eta_{j}\leq 2.8;\ \ \ \Delta R_{jj}\geq 0.4
\label{cutbasic}
\end{equation}

After carefully analyzing  the event samples of the signal and QCD background,
we apply the optimal selection cuts as
following.

\begin{itemize}

\item The four jets arising from QCD processes tend to be softer than that
of signal, thus we choose the first cuts as:
\begin{equation}
p_{T,j}\geq 60 GeV;\ \ \ \eta_{j}\leq 1.8;\ \ \ N_{j}=4.
\label{cut1}
\end{equation}

\item
After rearranging the four jets by the descending order of transverse
momentum, we apply the second cuts:
\begin{equation}
\begin{array}{rl}
\ p_{T,j_{1}}\geq 100 GeV;\ \ \ \ p_{T,j_{2}}\geq 80 GeV
\end{array}
\label{cut2}
\end{equation}

\item We pair the four jets into two groups by the criteria that
minimizes the quantity $|m_{j_{1}j_{2}}-m_{j_{3}j_{4}}|$ and relabel
the jets that $j_{1}$ and $j_{2}$ make up the first pair, and $j_{3}$ and $j_{4}$ make up the second pair .

\item Similar to the analysis conducted by ATLAS Collaboration \cite{Aad:2011yh}, we draw the $\Delta
R_{jj}$ of paired jets distribution in Fig. \ref{deltartev}.

\begin{figure}[htbp]
 \begin{center}
  \includegraphics[width=0.80\textwidth]
  {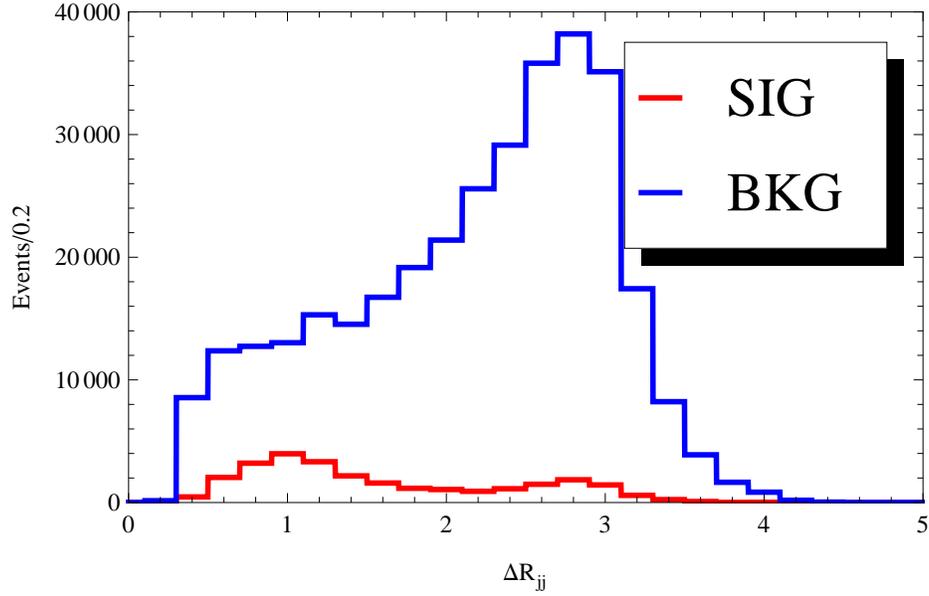}
  \end{center}
  \caption{$\Delta R_{jj}$ distributions of  signal and backgrounds at Tevatron. The luminosity is $10fb^{-1}$.}
  \label{deltartev}
\end{figure}

From the figure, we can see that the paired jets' $\Delta R_{jj}$  distributions of the signal and background are totally different. Thus we choose
the third cuts as:
\begin{equation}
\Delta R_{jj}\leq 1.6
\label{cut3}
\end{equation}

\item In order to show clearly how the proper $\Delta R_{jj}$ cut can greatly suppress the QCD background at the Tevatron,
we show the di-jet invariant mass distribution of the signal and background before and after the third cuts in
Figs. \ref{tevbefore} and \ref{tevafter}.

\begin{figure}[htbp]
\begin{center}
\includegraphics[width=0.80\textwidth]
{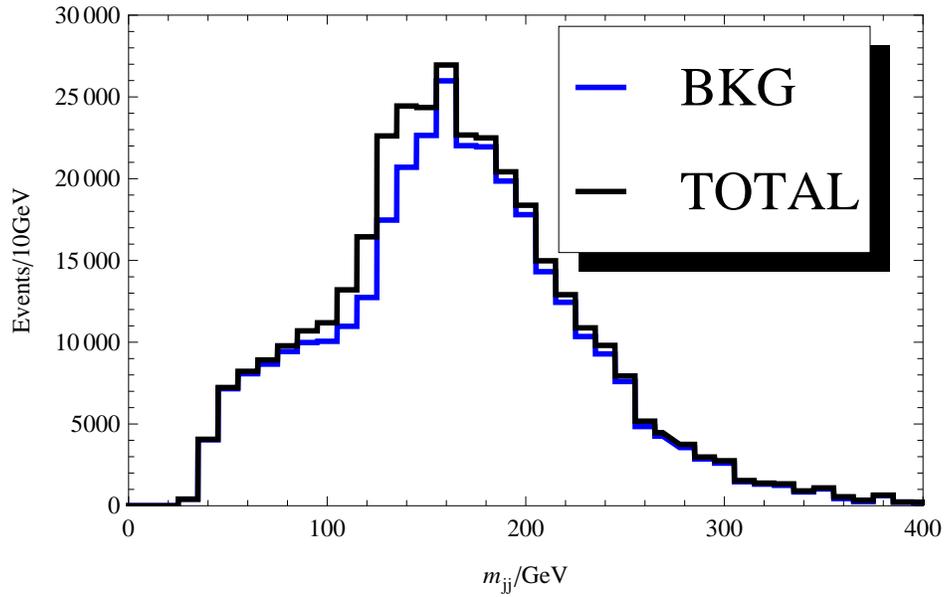}
\end{center}
\caption{Di-jet invariant mass distribution of signal and background at Tevatron {\em before} $\Delta R_{jj}$ cut. The luminosity is $10fb^{-1}$.}
\label{tevbefore}
\end{figure}

\begin{figure}[htbp]
 \begin{center}
  \includegraphics[width=0.70\textwidth]{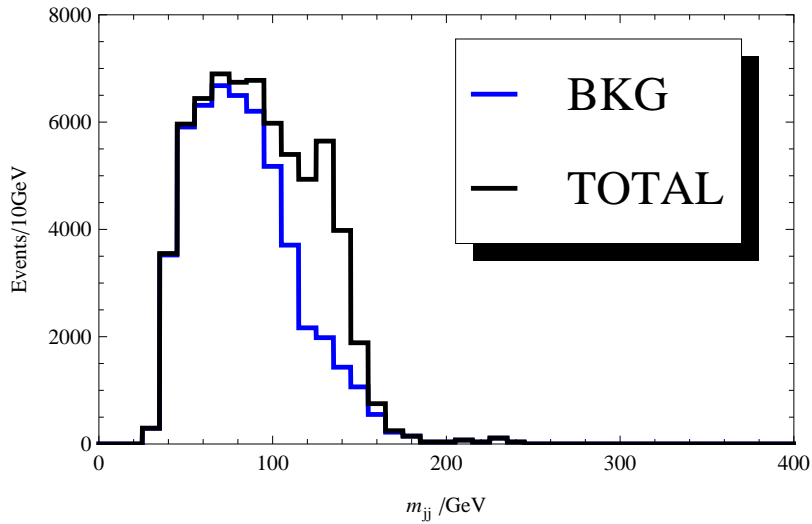}
  \end{center}
  \caption{ Di-jet invariant mass distribution of signal and background at Tevatron {\em after} $\Delta R_{jj}$ cut. The luminosity is $10fb^{-1}$.}
  \label{tevafter}
  \end{figure}

\item In order to isolate signal from backgrounds further, we apply the fourth cut, namely taking a mass window at the peak:
\begin{equation}
\Delta m_{jj} < 30\ GeV
\label{cut4}
\end{equation}

\end{itemize}

The signal and background event numbers vary after applying cuts conditions of Eqs. (\ref{cutbasic},\ref{cut1},\ref{cut2},\ref{cut3},\ref{cut4}), which are shown in Table \ref{tab1}.
In the real experiment, there are uncertainties for backgrounds estimation because of the experiments' equipments and other analysis processes.
For jets, the uncertainties are very large. In order to properly include such effects, we adopt the following formula to estimate significance of signal over background
\begin{equation}
\frac{S}{\sqrt{B}\bigoplus\gamma B}.
\end{equation}
Here the coefficient $\gamma$ is an empirical
parameter which is conservatively set to 0.3.

\begin{table}[htb]
\begin{center}
\caption{Event numbers for signal, QCD background, S/B and $\frac{S}{\sqrt{B}\bigoplus\gamma B}$
after applying cuts of Eqs. (\ref{cutbasic},\ref{cut1},\ref{cut2},\ref{cut3},\ref{cut4}) respectively
at Tevatron. The integrated luminosity is $10fb^{-1}$. }
\tabcolsep0.18in
\begin{tabular}{ccccccc}
\hline\hline Selection cuts & Signal & $\sigma_{Signal}$ & QCD background & $\sigma_{QCD}$ & $\frac{S}{B}$ & $\frac{S}{\sqrt{B}\bigoplus\gamma B}$ \\
\hline
 Basic cuts & 45300 & 4.53pb & 3670000 & 367pb & 0.01 & 0.04 \\
 1st cuts & 20964 & 2.10pb & 463007 & 46.3pb & 0.04 & 0.15 \\
 2nd cuts & 18681 & 1.87pb & 330080 & 33pb & 0.06 & 0.19 \\
 3rd cuts & 11592 & 1.16pb & 52187 & 5.22pb & 0.22 & 0.74 \\
 4th cuts & 8950  & 0.90pb & 5517 & 0.55pb & 1.62 & 5.40 \\
\hline\hline
\label{tab1}
\end{tabular}
\end{center}
\end{table}

From the table, we can see that the proper selection cuts can effectively reduce the QCD background from $3.6 \times 10^6$ to about $5.5 \times 10^3$, while
the signal from $4.5 \times 10^4$ to about $9 \times 10^3$. With already accumulated $10\ fb^{-1}$ data, Tevatron has an excellent chance to
discover such kind of new particle in the four jets events.

\subsection{$Z_C$ at LHC}

\begin{figure}[htbp]
 \begin{center}
  \includegraphics[width=0.8\textwidth]{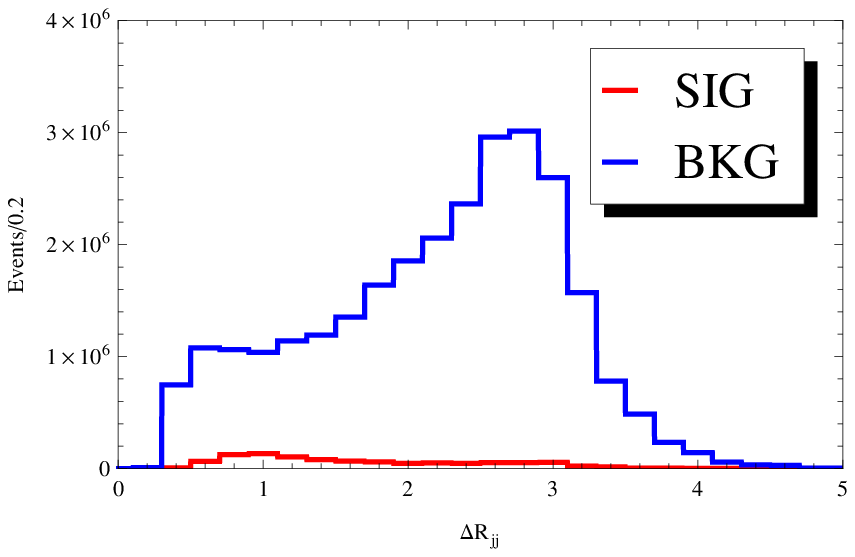}
  \end{center}
  \caption{ The $\Delta R_{jj}$ distribution of the signal and background at the LHC with $\sqrt{s}=7$ TeV. The luminosity is $10 fb^{-1}$. }
  \label{lhcdeltar}
\end{figure}

\begin{figure}[htbp]
 \begin{center}
  \includegraphics[width=0.8\textwidth]{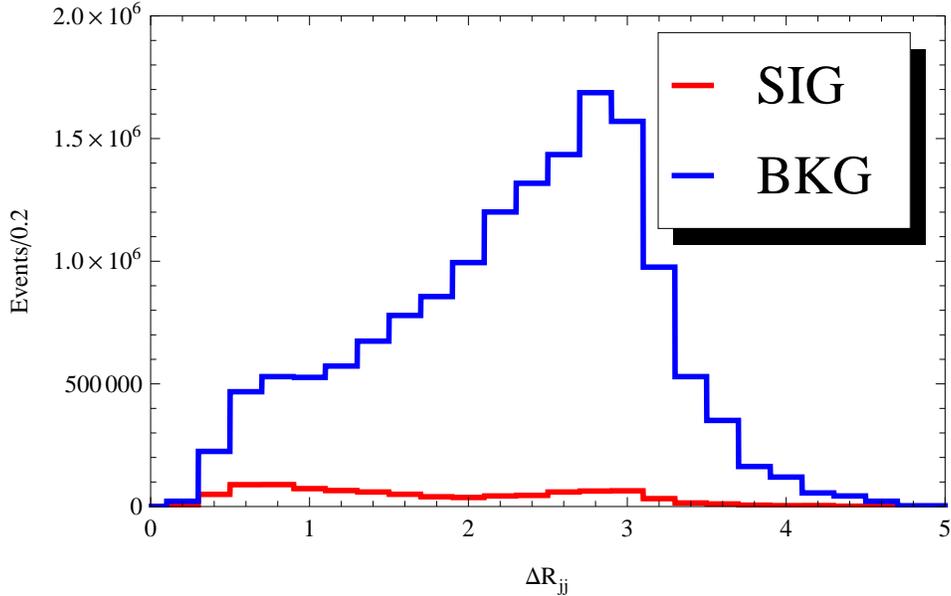}
  \end{center}
  \caption{ The $\Delta R_{jj}$ distribution of the signal and background at the LHC  with $\sqrt{s}=14$ TeV. The luminosity is $1fb^{-1}$. }
  \label{lhcdeltar14}
\end{figure}

Similarly we simulate the signal and background at the LHC with $\sqrt{s}=7$ and $14$ TeV respectively.
We find that the selection cuts for Tevatron are also suitable for LHC, so we use the same cuts in the analysis of LHC cases.
In Figs. \ref{lhcdeltar} and \ref{lhcdeltar14}, the $\Delta R_{jj}$ distributions of the signal and background with
$\sqrt{s}=7$ and $14$ TeV respectively are depicted. From figures, we can see the QCD background is much greater than that of signal. Though the signal is much larger than that at Tevatron, as also shown in Fig. \ref{zcpairlhc}, the QCD background at LHC becomes even larger.

\begin{figure}[htbp]
 \begin{center}
  \includegraphics[width=0.80\textwidth]{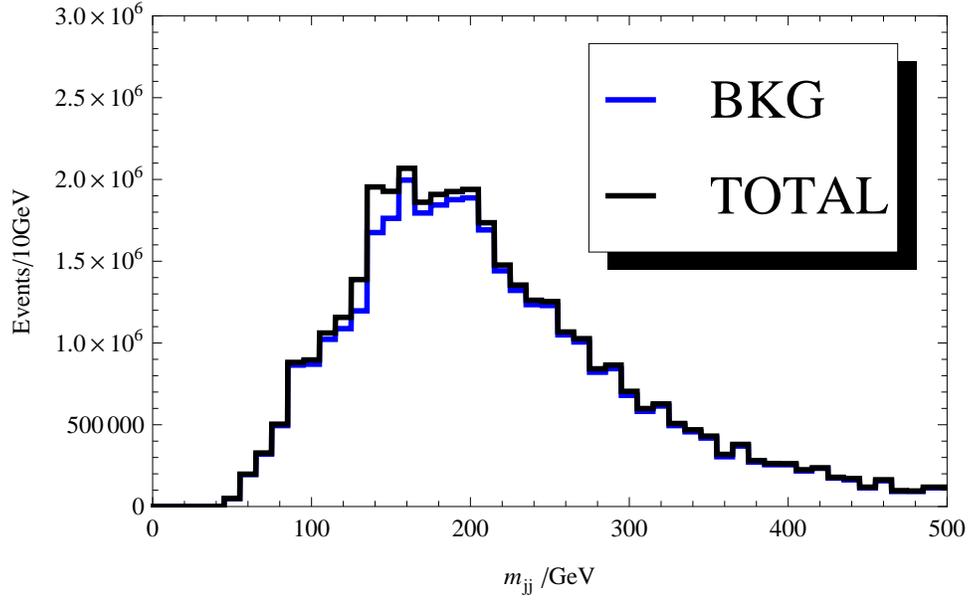}
  \end{center}
  \caption{Di-jet invariant mass distribution of signal and background at LHC with
   $\sqrt{s}=7$ TeV {\em before} $\Delta R_{jj}$ cut. The luminosity is $10fb^{-1}$.}
   \label{lhcbefore}
\end{figure}

\begin{figure}[htbp]
 \begin{center}
  \includegraphics[width=0.80\textwidth]{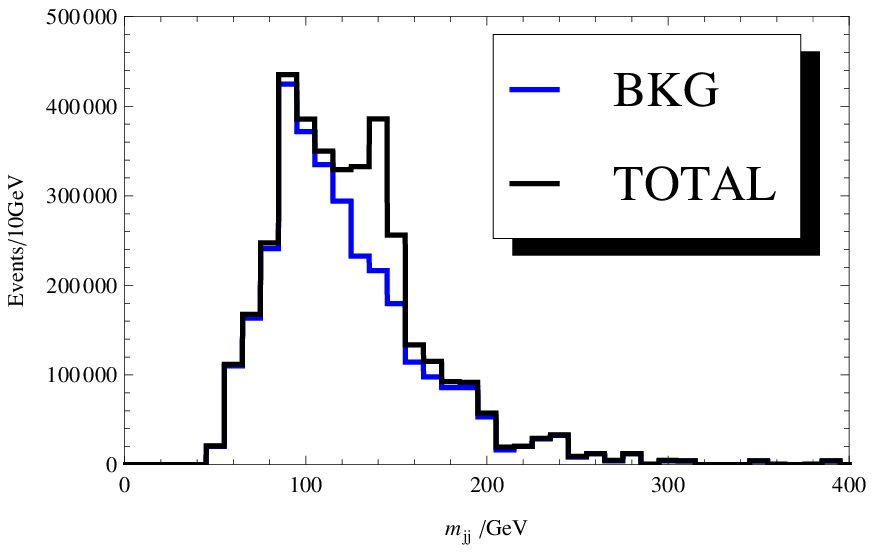}
  \end{center}
  \caption{ Di-jet invariant mass distribution of signal and background at LHC with
   $\sqrt{s}=7$ TeV {\em after} $\Delta R_{jj}$ cut. The luminosity is $10fb^{-1}$. }
  \label{lhcafter}
\end{figure}

\begin{figure}[htbp]
 \begin{center}
  \includegraphics[width=0.80\textwidth]{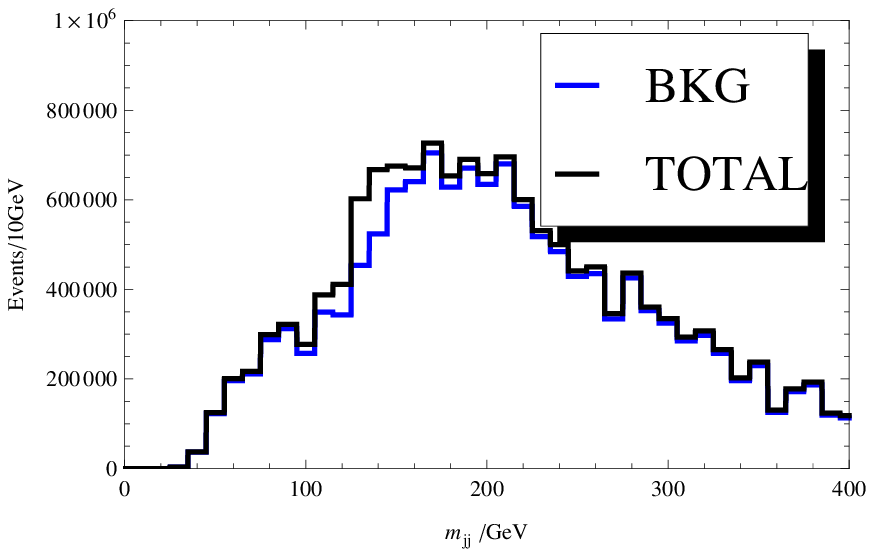}
  \end{center}
  \caption{Di-jet invariant mass distribution of signal and background at LHC  with
   $\sqrt{s}=14$ TeV {\em before} $\Delta R_{jj}$ cut. The luminosity is $1fb^{-1}$.}
   \label{lhcbefore14}
\end{figure}

\begin{figure}[htbp]
 \begin{center}
  \includegraphics[width=0.80\textwidth]{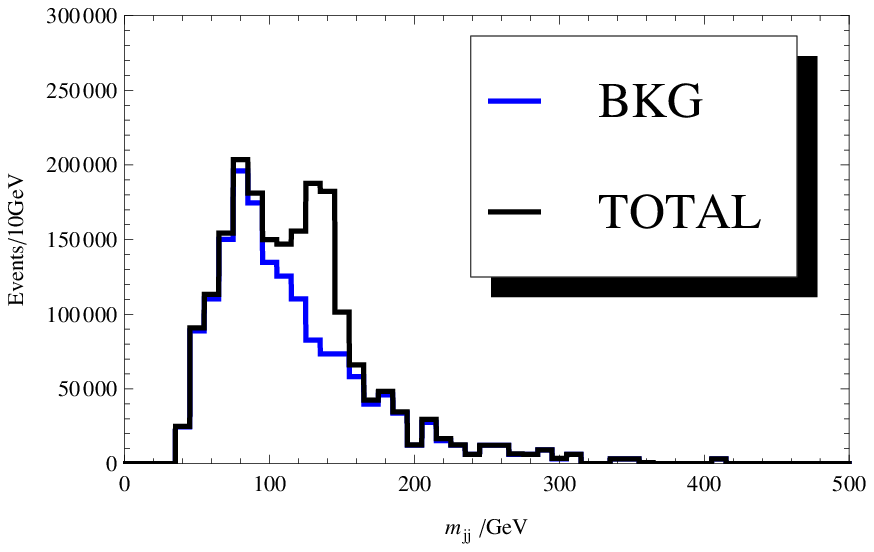}
  \end{center}
  \caption{Di-jet invariant mass distribution of signal and background at LHC  with
   $\sqrt{s}=14$ TeV {\em after} $\Delta R_{jj}$ cut. The luminosity is $1fb^{-1}$. }
  \label{lhcafter14}
\end{figure}

In order to show how the third cuts can suppress the QCD background
at the LHC, we show
the di-jet invariant mass distributions for signal and background
{\em before} and {\em after} the third cuts in Fig. \ref{lhcbefore} and Fig. \ref{lhcafter}
for $\sqrt{s}=7$ TeV with $10 fb^{-1}$ luminosity, as well as
Fig. \ref{lhcbefore14} and Fig. \ref{lhcafter14} for $\sqrt{s}=14$ TeV with $1 fb^{-1}$ luminosity.

After applying all the optimized cuts, we list the final results in Table \ref{tab2}
at the LHC with $\sqrt{s}=7$ and  $14$ TeV respectively.
From the table, we can see that LHC-14 with 1 $fb^{-1}$ data can
discover the new color-octet particle with large significance. Compared with Tevatron, due to the larger QCD background,
the capacity to discover the new particle at LHC is not better. Such results are not strange if one compares the capacity to
measure di-jet events of CDF/D0 and UA2.

\begin{table}[htb]
\begin{center}
\caption{Signal, background, S/B and $\frac{S}{\sqrt{B}\bigoplus\gamma B}$ at the LHC after applying all optimized cuts
of Eqs. (\ref{cutbasic},\ref{cut1},\ref{cut2},\ref{cut3},\ref{cut4}). }
\tabcolsep0.15in
\begin{tabular}{ccccccc}
\hline\hline$ LHC $ &Signal & $\sigma_{Signal}$ & QCD background & $\sigma_{QCD}$ & $\frac{S}{B}$ & $\frac{S}{\sqrt{B}\bigoplus\gamma B}$ \\
\hline
7TeV$\left(10fb^{-1}\right)$& 347710 & 34.77pb & 620690 & 62.07pb & 0.56 & 1.87\\
14TeV$\left(1fb^{-1}\right)$ & 300000 & 300pb & 265518 & 265.5pb & 1.13 & 3.77 \\
\hline\hline
\end{tabular}
\label{tab2}
\end{center}
\end{table}

%From the table above, we can see that the cut flow is also can suppress the huge background to make the signal exceed the background. So we can expect more luminosity on LHC or better cut flow can give us new physics. However, compare the figure of Tevatron and LHC,  we demonstrate that the Tevatron is more suitable to the search for multi-jet final states. We also suppose the result of LHC with $14TeV$ will be better, so we set the same cut flow on the signal and background on LHC, the event number we show in table 2. To compare and understand the result, we also show the result in figure 7.

%\begin{figure}[htbp]
% \begin{center}
% \includegraphics[width=0.45\textwidth]{soverb_lhc_before.eps}
%  \includegraphics[width=0.45\textwidth]{deltaR_signal_background_lhc.eps}
%    \includegraphics[width=0.8\textwidth]{soverb_lhc.eps}
%  \end{center}
%  \caption{Result on LHC with $\sqrt{s}=14TeV$ and $L=34pb^{-1}$. Above two figures shows the condition before $\Delta R_{jj}$ cut. From the figure, we choose $\Delta R_{jj}\leq 1.6$. Below figure shows the signal+ background and background events distribution with $m_{jj}$ after $\Delta R_{jj}$ cut. The size of each bin is 10GeV, the luminosity is $1fb^{-1}$ }
%\end{figure}

%From the figure of result of LHC, we can find the cut flow is better to suppress the huge background compared to the case of $7TeV$. At the same time, with the increasing of $\sqrt{s}$, the signal grows more illustrious than the background. As a result, the significance is more notable to help us to find the 4 jets signal. So we hope the future result on LHC.

\section{Conclusion and Discussion\label{conclusion}}

To summarize, we demonstrate in this paper that high energy hadron colliders, namely Tevatron and current running LHC,
can be utilized to investigate the O(100 GeV) physics beyond the standard model via the four jets events.
We simulated the signal and the corresponding background for the previously proposed new color-octet axial-vector $Z_C$ with a mass of 145 GeV, as indicated
by CDF di-jet bump. At the same time $Z_C$ can also accommodate the top quark forward-backward asymmetry measurements by Tevatron, which showed certain
deviation from the SM prediction. Our results showed that both Tevatron and LHC have the excellent chance to discover $Z_C$ through analyzing the four jets events
with accessible integrated luminosity and good control of QCD background.
Moreover we pointed out that Tevatron and LHC can play a role to identify the different production mechanisms, namely at LHC $Z_C$ pair production
comes from gluon-gluon contributions while at Tevatron the quark contributions are larger. In a sense Tevatron can be a better
collider to observe the multi-jet signal arising from quark annihilation.
On the other hand, with the higher energy and the integrated luminosity, LHC has the power to give the better result.

\section*{Acknowledgment}

This work was supported in part by the Natural Science Foundation
 of China (Nos. 11075003 and 11135003).

\bibliographystyle{h-physrev}
\bibliography{reference}

\end{document}